# A SYSTEMATIC REVIEW OF NATURAL LANGUAGE PROCESSING FOR KNOWLEDGE MANAGEMENT IN HEALTHCARE

[1]Ganga Prasad Basyal, [2]Bhaskar P Rimal and [1]David Zeng

[1]College of Business and Information Systems,
Dakota State University, SD, USA
[2]The Beacom College of Computer and Cyber Sciences,
Dakota State University, SD, USA

## ABSTRACT

*Driven by the visions of Data Science, recent years have seen a paradigm shift in Natural Language Processing (NLP). NLP has set the milestone in text processing and proved to be the preferred choice for researchers in the healthcare domain. The objective of this paper is to identify the potential of NLP, especially, how NLP is used to support the knowledge management process in the healthcare domain, making data a critical and trusted component in improving the health outcomes. This paper provides a comprehensive survey of the state-of-the-art NLP research with a particular focus on how knowledge is created, captured, shared, and applied in the healthcare domain. Our findings suggest, first, the techniques of NLP those supporting knowledge management extraction and knowledge capture processes in healthcare. Second, we propose a conceptual model for the knowledge extraction process through NLP. Finally, we discuss a set of issues, challenges, and proposed future research areas.*

## 1. INTRODUCTION

Use of Natural Language Processing (NLP) has grown in the last few years, where it is preferred along with other techniques such as visualization for managing a large amount of data [1]. Natural Language Processing in healthcare is applied to various tasks such as identification of clinical relationship inpatient records [2], disease identification [3], disease classification [4] and disease prediction [5].

NLP tools are used for the prediction of disease symptoms and behaviour traits, Study by [6] suggested that NLP can be utilized for the prediction of suicide behaviour by analyzing the clinical text Notes. The author states that Clinical notes (part of Electronic health records) can be very crucial in medical informatics research which often is not utilized up to full potential. Electronic Health records contain much useful information regarding the history of patient diagnosis, treatment, prescription, healthcare status but those are in the form of text (structured and unstructured) analyzing text data is very difficult and becomes more tedious in case of raw text which is highly unstructured, [2]. Information extraction from the data is of immense use as up to 18% of patient safety errors could have been reduced if proper information is available at the time of the decision-making process [7].

## 2. RESEARCH AGENDA

Natural Language Processing (NLP) has been utilized for knowledge management in the healthcare domain. NLP helped in gaining knowledge from a domain such as genetics under the field study of biomedical literature [8]. The author describes that genetic disorder can be related to the variation in individual genetics and can affect the medical condition or response to therapy. Gaining knowledge in the area will provide clinicians with research evidence that can be utilized for patient care.

The knowledge base in the healthcare domain is the primary source of data acquisition. These knowledge database exits in various forms, Electronic Health Records (EHR) is one of the major KM knowledge bases incorporating the patient information including patient visits, diagnosis and treatment course, diagnostic codes, patient lab reports, disease history, preventive action, clinician and nursing staff notes, etc. The advancement in EHR data management and the amalgamation of these data sources with Natural Language processing is facilitating medical information research. The data source can further be divided into two major availability sources, namely open-source and private sources. Open-source databases are those which are available for public viewing and usage. These databases can be in the form of online data provided by government agencies such as – National Health and nutrition examination Survey, the US department of agriculture data (USDA) [9]. Private sources, on the other hand, are much difficult to access as they require either permission or collaboration to access, these data sources can be in form of online information portal (i2b2 [2], patient information system), or private hospital database [10]. The Discovery of such KM data sources flourishes the common platform under which the data resides in tacit form can be converted into explicit form.

NLP tools are used for extracting the data from databases such as electronic medical records and clinical note documents [11]. Information extraction is very important in terms of obtaining complete information through raw data. Extracting information through the manually annotated document is used for identifying the phenotypic information related to inflammatory bowel disease [11]. Knowledge extraction from the large set of documents such as EHR required certain extraction tools that are capable of identifying the relationship between the text and extract the important information in form of features for higher-level understanding [12], for example extracting the particular medical terminology such as drug reaction from the whole EHR text can be done by using some of the NLP tools. A study by [13] suggests and hybrid model algorithm capable of detecting the discrepancy in medication. The author suggests an algorithm comprising of Machine learning and Natural language processing system to provide accurate detection based on gold-standards. The accuracy shown by the model in entity detection is above 95% which is promising but the detection in medication had less accuracy i.e. 65% approx. which implies that the model still needs to be tuned properly to be viable for commercial application. Discharge summaries had been used as a source of data and analysis performed on overview notes those are part of lengthy discharge notes of the pediatric patient database from Complex care medical home program based in Cincinnati children's hospital and medical center. The author used the Conditional random field (CRF) and Universal medical language system (UMLS) for the knowledge extraction system enabling feature creation, both for entity detection and medication discrepancy matching.

Knowledge gained in the domain should also be assessed in terms of quality of knowledge base created, as this is one of the major challenges faced by knowledge management systems. A study by [14] addresses this challenge pertaining to the healthcare field knowledge base by assessing the quality of psychotherapy note texts their implication focused on Post-treatment stress syndrome (PTSD). The results show that more than half of the service had been coded wrongly under psychotherapy which is eye-opening knowledge gained from research.

## 3. LITERATURE REVIEW

Studies related to information extraction from the EHR had been done in the past [11] where authors attempted to assess the role of the NLP system for extracting the information in the form of phenotypic data. The process is manual as it requires to create reference standard and manual annotations from clinical text notes [11]. Another study where the author attempted to embrace the medical knowledge as a key to clinical application is identified, study shows that medical knowledge updated time to time and therefore much of knowledge is stacked down in medical literature, medical knowledge is evidence-based [15] and needs empirical literature to capture information and extract knowledge out of the medical literature database.

Another study extracting the information through Electronic Health Records had been done by [16]. This study used General Architecture for text engineering (GATE) – the NLP system for knowledge extraction from Clinical records interactive search (CRIS) to evaluate the antipsychotic polypharmacy data. Results suggest that having high precision is good for the current data but the low recall is the result of a trade-off between the decision of precision and recall and High precision suggests this tool is good for polypharmacy detection from mental health EHR's.

Studies suggesting extraction of information regarding polycystic ovary syndrome [17] and influenza-like illness symptoms using a rule-based classifier [18] had been done. Research suggests that the algorithm was able to identify a problem area of clinical narrative classification as only 9.2% of all detected ILI enter in any month but future research is needed to increase the positive predictive value with another dataset.

Information extraction from patient records could provide very valuable information which is hidden and mostly untraced through the normal skimming or overview, natural language processing provides the automatic recognition of clinically important concepts and the identification of the relationship between the patients records to extract the knowledge in form of evidence which can be used for medical decision support and application, following relationship had been identified through the model proposed in [2]. The database used for information extraction was the i2b2 server and 826 patient records had been studied in research. The author states that this model could be used for information capture and discovery task in medical informatics. For information extraction stemming, stop-words elimination had been used and MetaMap is used for feature extraction under the NLP.

Information extraction from medical journal and publication is used to be of important use, a combination of the NLP and machine learning yielded very promising results in research conducted by [4]. These concepts had been extracted through the Unified medical language system (UMLS) database utilizing NLP system applications such as bag-of-words. As per the research, NLP concepts such as Biomedical concept representation is used for biomedical concept identification. The author selected a journal research article published in pub-med for extracting the information from text articles.

NLP facilitates knowledge management through the detection of disease symptoms from clinical notes [3]. These notes contain free text information from emergency department admission and discharge, details related to patient allergy, medication, previous visits had been mentioned in the clinical records. This study involves a large sample population i.e. over 40,000 and two major hospitals had participated in the study. This research was focused on the identification of common symptoms responsible for the transferability of influenza cases. The knowledge extraction process involves natural language processing system – NLP parser (pattern matching and deduction of rules) this system is capable of retrieving the information through implementing the rules created and provide the matching results.

The knowledge capture process is iteration-based, therefore, knowledge captured is always updated and changed according to the update in the domain but there is always a chance of error in the knowledge update process. This can be avoided through validation and automation of knowledge base through software process but in past EHR had been entered manually in the systems which are prone and have errors. Research conducted by [19] suggests the NLP system proposed as a negation algorithm capable of extracting symptoms reported by the patient and matching it with documented medical records.

A study by [20] aims at identifying patient exposure towards cardiovascular risk and contraindication with the help of the NLP of EMR. The author attempts to identify the aspirin usage of the patient from EMR using NLP and states that it could be a risk factor for cardiovascular disease. The proposed algorithm achieved the sensitivity of 81% compared to manual results of 84% shows that the automated process is almost equally effective as manual work. The study had limitations such as the text could only be used specifically for over the counter medication and more information is a need in part from self-report by patient, pharmacy transaction report to have accurate.

Research by [21] suggest that Natural language processing system can be used as the knowledge tool for understanding whether there is any association exists between HIV and smoking history, the association could be direct or independent. The study suggests that Less controlled HIV had fewer chances of quitting smoking even less than the chances with mental disorder to quitting smoking and Smoking found to be more prevalent in HIV-infected patients as compared to the normal sample.

The study had been done for early recognition of diseases, such as research by [22] this research provides support to KM through retrieving the 66 individual buckets (e.g., – "Paresis" had 14 different terms in ICD 9 code which then aggregated into one bucket and other like Fatigue, vision, weak, headache tingling, etc.) from Electronic health records of patients sample acquired through Columbia University Medical Center (CUMC) using natural language processing system to query based on the knowledge base of ICD-9 standards. This research is based on model-based on the model trained on the MS enriched patient dataset, the classifier was able to identify 40% of the population going to have MS before actually, it happened with AUC (0.71), sensitivity (40%) and specificity (97%). The author state one limitation as classifier was tested on local samples which is biased with the Hispanic population along with elder age and largely female patient.

Embedding the medical concepts for higher knowledge representation had been addressed in a study by [23] where the word2vec (NLP system) model is used on novel learning scheme capable of integrating the clinical notes with diagnostic code and laboratory codes. The result shows that model was successful in predicting the diagnosis codes based on the set of clinical notes and diagnosis codes and 4 Physicians testify that proposed model is far better than baseline model in predicting the phenotypes for 3 out of 6 diseases and for the rest 3 case they had split opinion but all agreed that the proposed model had higher relevance of phenotype. The only limitation the study had was words having very less or rare frequency are not included during pre-processing. The NLP system had been adopted for modelling disease severity [24] and modelling and extracting variability [25] in using Electronic health records.

Knowledge extraction for disease identification such as methicillin-resistant staphylococcus aureus (MRSA) [26] and adverse drug reaction [27] had been implemented using the NLP systems such general architecture for text engineering (GATE) and unstructured information management architecture (UIMMA). The studies suggest that the NLP system helps in advance identification of potential ADE from EHR and implementing the NLP system can be promising for the identification of pathogen surveillance data.

The research gap is identified through studies conducted in past related to information extraction that has been related to either one disease or symptoms [11], [3] the purview of the research should be holistic and require a broad focus which encompasses higher representation of knowledge and the information extraction from NLP system supported to knowledge management in healthcare.

Identification of the NLP system as a support system for knowledge capture and extraction in most of the research had been absent thereby creating a research gap, importance of medical knowledge had been advocated in some studies such as [15] but failed to establish the relationship between knowledge management and healthcare in particular. KMS support the information flow through the extraction of information from medical databases through NLP.

## 4. RESEARCH FRAMEWORK

Knowledge is the highest representation of the information gained from the raw data available. The data in its basic form does not contain any direct information. Thus, tools are required to analyze the data for extracting the information out of it. This continuous flow and assimilation of information create a knowledge-base, which is the highest level in the hierarchy.

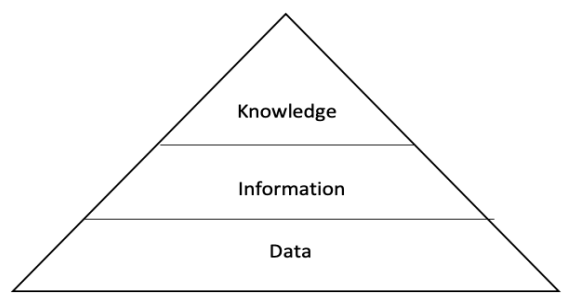

Figure 1. Hierarchy of knowledge

Knowledge is defined as the justified belief that increases an entity's capacity for effective action [28]. Knowledge management includes the KM process, which enables the smooth transition of knowledge. Two major studies in the field of knowledge management propose the four main steps. The KM process and the potential role of IT [28] incorporate four steps in the process, namely, 1) knowledge creation, 2) knowledge storage/retrieval, 3) knowledge transfer, and 4) knowledge application. On the other hand, the KM process stated by [29] also consists of four parts: 1) knowledge discovery, 2) knowledge capture, 3) knowledge sharing, and 4) knowledge application. We adopt the KM process proposed by [28] for our research methodology.

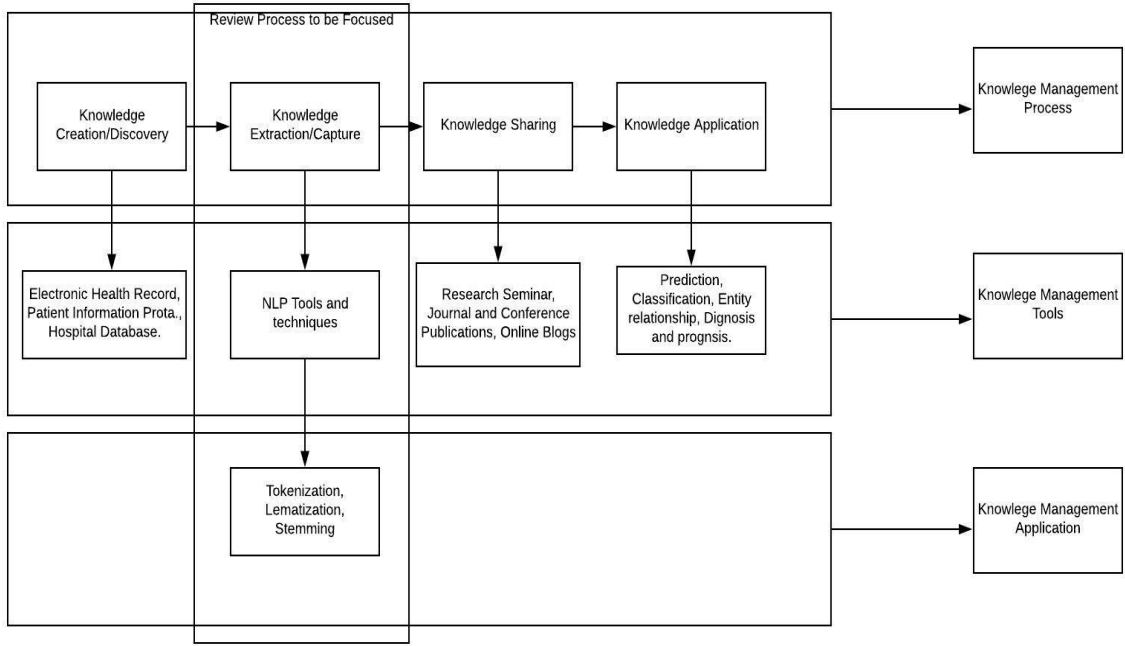

Figure 2. Research methodology framework

Our research focus is on the second process described by Alavi et al. and Beccerra et al. Both authors proposed a similar approach to the knowledge extraction one had named is as knowledge capture and others said it knowledge extraction. Knowledge created is forgotten over time, which may be due to less or no use thus this memory [28] should be retrieved before it elapses and fades away. Our review study would provide a comprehensive review of the knowledge management extraction process supported by the NLP focusing on the healthcare domain.

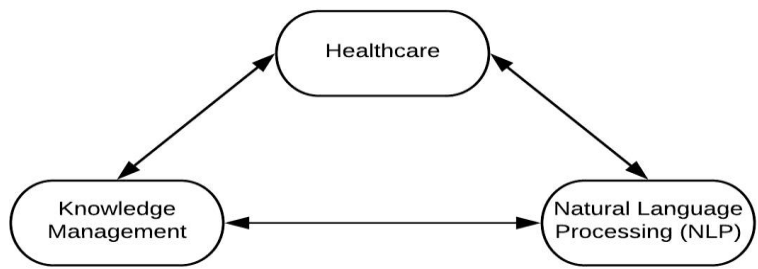

Figure 3. Knowledge flow in the healthcare domain

The NLP supported by knowledge management extracts/captures the information. The knowledge management helps in the integration of cross-domain information flow. The healthcare domain is rich in data, where information is needed for every small and big decision [7]. This information can be extracted with the help of natural language processing. Fig. 3 illustrates how knowledge management supports the NLP in extracting the information from the Healthcare domain.

For selection and screening of research article for literature review, we referred the model proposed by [30] this model had been considered as very efficient as it provides the step by step checklist for the inclusion and exclusion criteria. The model is bifurcated in four steps 1) identification, 2) screening, 3) eligibility, and 4) included.

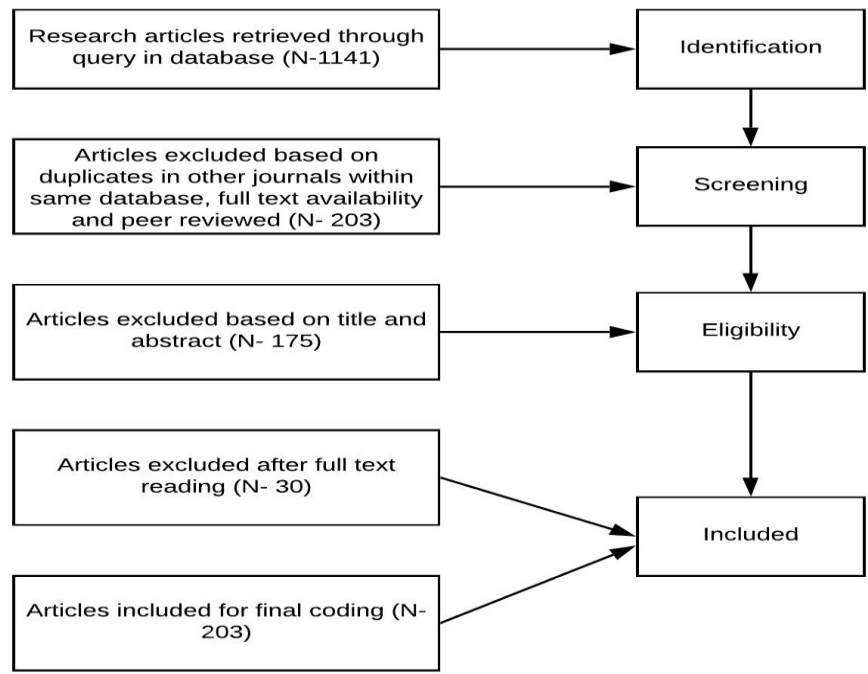

Figure 4. Article selection and screening chart ([30])

We query the EBSCO database which is considered to be one of the largest database sources for free research articles containing all major journals such as BMC, PLOS ONE, IEE, etc. During the first step, the resultant query provided us 1141 research article which we filtered further in the second step of screening. In screening, we applied 3 exclusion criteria: 1) removing duplicates, 2) peer-reviewed journal articles, and 3) full-text availability of the articles. The number of articles after the second step is 203. The third step consists of filtering articles based on the relevance of title and abstract with our research objective. The final filter applied was based on the full reading (skimming) of the article aligning with our research objective.

## 5. FINDINGS AND LIMITATIONS

The NLP system supports knowledge management process of information capture and extraction through specialized tools such as concept extraction based text analysis system (CETAS) for concept identification [5] and extraction in line with other application such as named entity recognition (NER) and conditional random fields (CRF) [2] had also been implemented. The NLP supports knowledge management for validation or evaluation of knowledge captured vs. knowledge documented [19]. Maker and checker in the medical information system such EHR are human and human is more vulnerable towards mistakes this problem is addressed and solved through NLP negation algorithm which compares the patient-reported symptoms with information documented.

Knowledge management process of externalization is also supported by the NLP through converting the tacit knowledge stored in medical database systems such as the unified medical language systems and converting it to explicit information [4], [13]. The NLP supports knowledge management in information capture and extraction through the NLP systems such as MedLEE proposed by [15], the general architecture for text engineering (GATE) by [16], and [27], the Word2Vec by [23], the conditional random field by [14], the Unstructured Information Management Architecture (UIMMA) [26].

We identified that majority of the research had adopted some kind of quantitative measures to compare, contrast, evaluate their research with either previously conducted similar research models or evaluating research with the same data having two splits – training and test data. [4], or to establish a new milestone with completely new methods or a hybrid model [13]. The research studies had compared their performance of two major evaluating measure – precision-recall matrix [13] and sensitivity and specificity matrix [20]. Both the matrix has a similar concept of accuracy calculation but the representing terminology is different. It depends upon the research choice which one they want to adopt.

Research studies such as [19] [13],[15], [21] suggest that the performance of the proposed model [15] had performed well providing the precision of 86% in identifying the association of phenotypes in different patient reports. 86% accuracy achieved in drug symptom and 77% for disease symptoms association and some studies even above 90% [14], [20].

We identify that many of the research studies had utilized the data to trace the prevalence of the disease [21], [11] study by Toyabe et al. provide the finding related to smoking habit stating that regular smoking percentage (33%) is less than the percentage of patients who ever smoked (47%) in life which shows that many of the patients tried and stopped the habit of smoking at some period in their life. Researches reveal that medical data is capable of disease management based on learning from the empirical or historical dataset, [23] states that their model automatically traced the diagnosis codes from the patients' EHR which can be useful for patients' prognosis and records management.

We identified that data quality is one of the barriers to achieving the set accuracy targets. The major problem with data quality is missing data. Data acquired from an open source such as online platform is low in term of quality and need much more time in cleaning and pre-pressing to get acceptable quality. Another limitation is the sample size of data which tends to influence the accuracy of results [20], [24], [18], [31], [14] some of the studies even having sample sizes less than 1500 to as low as 221. This states that a limited number of observations hampers the performance of models which ultimately results in low accuracy.

# 6. CONCLUSION AND FUTURE AREA

Our systematic review provides a comprehensive overview of the knowledge management process, which includes knowledge extraction/capture as a sub-process supported by the NLP system focused on the healthcare domain. We conclude that knowledge management plays a very important role in the extraction of healthcare domain knowledge. This domain knowledge is further applied in the field as a knowledge system for increasing the efficiency of the designed model, which ultimately leads towards evaluating the domain challenges and serving as a decision-making tool to overcome such challenges.

Our review shows that clinical notes and patient records contain very useful and rich information but in an unstructured form which required explicit process and tools for extracting knowledge out of them. Studies had suggested ways to extract that information but had a limitation of ignoring medical data language, due to which key medical terms such as tightness, senses, anginal symptoms, and dyspnoeic had been left, resulting in low accuracy. Further research is needed in terms of model hyper tuning, which can be addressed as shown in the research framework. Future research areas could be the assessment of physician and clinical notes through information system tools in the healthcare domain.

## AUTHORS

**Ganga Prasad Basyal** is pursuing his Ph.D. in Information Systems with a specialization in decision support and analytics. He received his MBA from D.A.V.V University, India and Master in Information Systems (MSIS) from Dakota State University (DSU), Madison, SD. Currently, he is a teaching assistant at the College of Business and Information Systems, at DSU. His research area includes healthcare analytics with the inclusion of data mining and deep learning technologies.

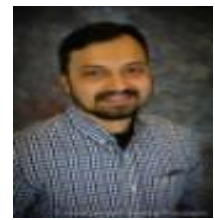

**Bhaskar Prasad Rimal** (Senior Member, IEEE) is an assistant professor of Computer and Cyber Sciences at Dakota State University, Madison. He received the Ph.D. degree in telecommunications engineering from the University of Quebec, Canada. He was a Postdoctoral Fellow with the Department of Electrical and Computer Engineering, University of New Mexico, Albuquerque. He was a Visiting Scholar with the Department of Computer Science, Carnegie Mellon University (CMU), Pittsburgh. He was a Visiting Assistant Professor and Academic Specialist at the Department of Computer Science at Tennessee Tech University, Cookeville. He has been serving as an Associate Editor of the IEEE Access, Associate Technical Editor of the IEEE Communications Magazine, Associate Editor of the EURASIP Journal on Wireless Communications and Networking, and an Editor of the Internet Technology Letters.

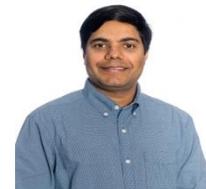

**David Zeng** is an assistant professor of Information Systems at Dakota State University, Madison, SD. David received his MBA and Ph.D. in MIS from the University of California, Irvine. His research focuses on the economics of IT-enabled services, the impact of IT-related factors on competition, market, and adoption of innovative services, and application of data analytic techniques in the healthcare industry. His work has been published in peer-reviewed journals and won best-paper awards in conferences. David taught undergraduate, MBA, and MS courses in China and South Korea, in areas related to management information systems, supply chain and operations management, and data analytics.

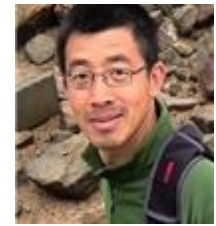